**RESEARCH ARTICLE**

# Quantum Blockchain Based on Dimensional Lifting Generalized Gram-Schmidt Procedure

**KUMAR NILESH**[1] **AND PRASANTA K. PANIGRAHI**[2]
[1]Department of Mathematics and Statistics, Indian Institute of Science Education and Research Kolkata, Kolkata 741246, India
[2]Department of Physical Sciences, Indian Institute of Science Education and Research Kolkata, Kolkata 741246, India

Corresponding author: Kumar Nilesh (kumarnilesh5198@gmail.com)

The work of Prasanta K. Panigrahi was supported by the Department of Science and Technology (DST), India, under Grant DST/ICPS/QuEST/Theme-1/2019/6.

**ABSTRACT** The advancement of quantum computers undermines the security of classical blockchain, necessitating either a post-quantum upgrade of the existing architecture or creation of an inherently quantum blockchain. Here we propose a practically realizable model of a fully quantum blockchain based on a generalized Gram-Schmidt procedure utilizing dimensional lifting. In this model, information of transactions stored in a multi-qubit state are subsequently encoded using the generalized Gram-Schmidt process. The chain is generated as a result of the reliance of orthogonalized state on the sequence of states preceding it. Various forking scenarios and their countermeasures are considered for the proposed model. It is shown to be secure even against quantum computing attacks using the no-cloning theorem and non-democratic nature of Generalized Gram-Schmidt orthogonalization. Finally, we outline a framework for a quantum token built on the same architecture as our blockchain.

## I. INTRODUCTION

The initial concept of Distributed Ledger Technology (DLT) can be traced back to the Byzantine Generals Problem [1], before Blockchain and Bitcoin were introduced. The notion of Blockchain that we know today, i.e., a cryptographically linked decentralized and distributed database that forms a chain using timestamped electronic data, was first pioneered by Haber and Stornetta [2] and Bayer *et al.* [3]. The accountability and transparency of transactions is a key feature of a blockchain, which makes them appealing for a wide range of applications [4] such as in healthcare industries [5], government organizations [6], finance [7], et cetera. The essential components of a blockchain network are [8]: (1) Nodes, which is a form of electronic device that keeps the network running by maintaining local copies of the blockchain; (2) Blocks, each of which has a cryptographic hash of itself and of the previous block, as well as a timestamp and transaction data. The timestamp establishes that the transaction data existed at the time of the block's publication; (3) Miners, who create new blocks on the chain through a process called mining. One of the most important concepts in blockchain technology is decentralization, in which the authenticity is confirmed by the community of nodes. Further, it is a distributed ledger in which records are in a shared form, ensuring no single organization controls the entire system and also eliminating the use of intermediaries. Even though every node has its own copy of the blockchain, they are transparent, and every action in the ledger can be easily viewed and verified. Each block also contains information from the previous block, thus forming a linear chain. This makes blockchains resistant to data modification because data in a particular block cannot be modified retroactively without affecting all subsequent blocks. That is, if a hacker attempts to modify a block, the hash will change, affecting all subsequent blocks.

The debut of Bitcoin [9] in 2008 signaled the start of a new era in blockchain technology. Bitcoin being a self-sufficient, anarchic system, requires the authority of no single central organization. However, the data which carries monetary value is controlled and authenticated by each peer in the distributed network. The democratic aspect of the network is determined here by the amount of processing capacity of each node rather than the number of members. Validation of any transaction is directly linked with the computational power one carries.

The associate editor coordinating the review of this manuscript and approving it for publication was Chi-Yuan Chen.

Miners who validate the legitimacy of Bitcoin transactions need to attach a valid SHA256 hash to the block's header. In return for their computational work, miners are rewarded with Bitcoin (native token), which is the mechanism to introduce new bitcoins into circulation. In addition to its own hash, a block keeps the hash of the block before it, resulting in the formation of a chain. At any instant, nodes will follow the longest chain in the network, considering that most cumulative work is done in that chain and discarding any other chain from their local repository. Blockchains like Bitcoin also allow for the creation and execution of smart contracts, which are on-chain automated programs with multi-signature and hashed timelock [10]. As the market capitalization of Bitcoin grows, new cryptocurrencies such as Ethereum, Ripple, Alastria, and others have begun to appear [8]. Despite the fact that blockchain and cryptocurrencies have been around for a decade, the distributed technology is still in its infancy and is finding new applications in every facet of digital information. The rising expense of mining, on the other hand, is encouraging investors to consider alternate methods for validation. Further, confirmation of each transaction takes a lot of time, making transactions very slow.

A blockchain relies on two asymmetrical cryptography: cryptographic hash functions and digital signatures [11]. Security based on mathematical complexity is not unique to blockchain. Most of the current cryptography is based on certain kinds of mathematical encryption that are difficult to solve with current computational capabilities [12]. Therefore, one of the most intriguing applications of quantum computers is breaking the mathematical difficulty that forms the base for currently used cryptography [13]. With the discovery of the Shor's algorithm [14], it was apparent that a sufficiently powerful quantum computer could break the mathematical difficulty of asymmetrical cryptography. Quantum computers can speed up the mining process and crack the SHA256 hash algorithm used by the Bitcoin network, thus making it vulnerable to 51% attack [15]. Quantum computers could also completely destroy Bitcoin's classical signature. Additionally, quantum computers will be able to modify the data contained in a block without altering the hashing function. When the hash is unchanged, the chain will appear undisturbed and intact. This seems to make Blockchain completely pointless and useless in the post-quantum world.

In the domain of key distribution, we have already seen the development of information-theoretically secure quantum cryptography protocols [16], [17], [18]. And unless quantum technologies are integrated into blockchain technology, the current classical blockchain may fail [19]. While several attempts have been made to incorporate post-quantum cryptography, they do not provide unconditional security [20], [21], [22]. In order to assure authentication in the post-quantum world is to use quantum computations, which provide (unconditional) information-theoretic security based on presumably unbreakable laws of quantum physics instead of some mathematical complexity. And use quantum technologies to quantize the network, which keeps on evolving. Along with the security and robustness, the Quantum Blockchain protocol has several other advantages over classical blockchain, including immediate verification of transactions and less resource consumption in mining.

Threat from quantum computation inspired many recent studies in the conceptual design of quantum blockchain. Here the data are encoded in quantum states, which are then converted into a quantum block. Jogenfors, in 2016 simply introduced a quantum bitcoin scheme for transaction systems [23]. His scheme, however, was ineffective. Afterward, Rajan and Visser [24] using entanglement in time gave a conceptual scheme for quantum blockchain. It lacked many details and security analysis. Furthermore, implementing it on a large scale is difficult. More recently, quantum blockchain based on weighted hypergraphs was proposed [25]. Although the concept was interesting, but the consensus was incomplete, and there was no way to recover back the encoded information. As far as we can tell, the research and development of quantum blockchain is still in its early stage, and there are numerous issues that need to be addressed.

In this article, we propose a quantum blockchain protocol based on dimensional lifting generalized quantum Gram-Schmidt procedure. The construction of each block in the network is done by applying a non-democratic orthogonalization procedure to multi-qubit state vectors. The proposed scheme requires fewer quantum capabilities making it practically realizable even in a small-scale quantum network.

Section 2 introduces the orthogonalization method used in our protocol. The construction of our quantum blockchain protocol is then described step by step in sections 3 and 4. Possible forking scenarios and countermeasures are discussed in Section 5. Section 6 contains a security analysis that demonstrates robustness not only against currently known quantum computing attacks but also against those that may be discovered in the future, potentially making post-quantum cryptographic schemes vulnerable. Finally, section 7 describes a quantum token based on our proposed quantum blockchain.

## II. GENERALIZED QUANTUM GRAM-SCHMIDT ORTHOGONALIZATION

The original Gram-Schmidt orthogonalization procedure that constructs orthogonal states from an ordered set of linearly independent states is well-known [26]. This is a useful method, which can be used to produce an orthonormal basis set $|w_1\rangle, \ldots, |w_n\rangle$ from an ordered set of states $|v_1\rangle, \ldots, |v_n\rangle$ in the Hilbert space **H**. We can achieve this by defining $|w_1\rangle \equiv |v_1\rangle / \| |v_1\rangle \|$, and for $1 \leq k \leq n-1$ we define $|w_{k+1}\rangle$ inductively by

$$|w_{k+1}\rangle \equiv \frac{|v_{k+1}\rangle - \sum_{i=1}^{k} \langle w_i | v_{k+1} \rangle |w_i\rangle}{\| |v_{k+1}\rangle - \sum_{i=1}^{k} \langle w_i | v_{k+1} \rangle |w_i\rangle \|}$$

A simple example of this method in the context of qubits can be found in Appendix A. Several modifications have been proposed to enhance its efficiency and stability for specific scenarios where the original version might not be suitable [27].

In this paper, we will make use of Havlicek and Svozil's generalization of Gram-Schmidt orthogonalization, which employs dimensional lifting [28]. The constraint of linear independence on the initial states is relaxed in this generalization by extending the dimension of the original Hilbert Space. As a result, it broadens the range of states for which the Gram-Schmidt orthogonalization can be used. It calculates each component of the lifted dimension one by one in a recursive manner, and at each step, there is only one unknown scalar to be computed and thus provides more accurate results than the traditional Gram-Schmidt. The end result is demonstrably orthogonal and projecting along the newly added dimensions returns the initial states. Therefore, the information encoded in the initial states is preserved in the orthogonalized outcome.

Here we will give an alternative proof for this generalized Gram-Schmidt orthogonalization with the help of matrices.

*Theorem 1:* Let $v_1, \ldots, v_m$ be an arbitrary finite set of states (not necessarily linearly independent) in a Hilbert space $\mathbf{H} = \mathbb{C}^n$. Then there exists a set of orthogonal states $w_1, \ldots, w_m$ in the extended Hilbert space $\mathbf{H}' = \mathbb{C}^{n+m}$ and a surjective partial isometry $P : \mathbf{H}' \to \mathbf{H}$ such that $Pw_i = v_i$ for all $i = 1, \ldots, k$.

*Proof:* Let our initial states be $v_1, v_2, \cdots, v_m$, they might not be linearly independent. And $\mathbf{H} = \mathbb{C}^n$ represents the Hilbert space. Using the initial states as columns, let us construct a matrix

$$M = \begin{pmatrix} v_1 & \cdots & v_m \end{pmatrix} \quad \in \mathbb{C}^{n \times m} \tag{1}$$

Now if we define a new matrix as $A := r^{-1}M$ for any fixed $r > \|M\|_2$, we get $\|A\|_2 < 1$. With this matrix $A$, we define a new matrix $B$ as

$$B = \left(I_m - A^T A\right)^{1/2} \quad \in \mathbb{C}^{m \times m} \tag{2}$$

where $A^T$ is transpose of matrix $A$ and $I_m$ is $m \times m$ identity matrix. The existence of $B$ can be easily verified as $I_m - A^T A$ is a positive definite matrix. Now observing that

$$\begin{pmatrix} A^T & B^T \end{pmatrix} \begin{pmatrix} A \\ B \end{pmatrix} = A^T A + B^T B = I_m \tag{3}$$

implies $\begin{pmatrix} A \\ B \end{pmatrix}$ has $m$ orthonormal columns and thus can be extended to a orthogonal matrix $N$, forming the first $m$ columns of $N$.

Therefore, it is easy to notice that every (potentially rectangular) matrix $M$ can be extended to a nonzero scalar multiple of an orthogonal block matrix

$$N := \begin{pmatrix} M & * \\ * & * \end{pmatrix}$$

Further, the states $w_j$ forming the columns of $N$ are not only mutually orthogonal but also have equal norm $r$. The advantage of this technique is that we can recover the initial states by projecting along the additional dimensions. For $i = 1, 2, \ldots, m$, consider the following map

$$P = \begin{pmatrix} I_n & 0_{n \times m} \end{pmatrix} M \quad \in \mathbb{C}^{n \times (n+m)} \tag{4}$$

Then $P$ becomes a surjective partial isometry, defined by $P(w_j) = v_j$, which gives back our initial states. □

Gram-Schmidt orthogonalization is easy to implement in Quantum Computers as we already have Quantum circuits for Gram-Schmidt procedure [29], [30], taking input: linearly independent states $\mathbf{b}_1, \mathbf{b}_2, \ldots, \mathbf{b}_n \in \mathbb{R}^n$ and giving output: mutually orthogonal basis states $\mathbf{b}_1^*, \mathbf{b}_2^*, \ldots, \mathbf{b}_n^*$ and a transformation operator $\mathbf{M}$.

## III. QUANTUM BLOCKCHAIN

Taking into account the protocol's overall structure, which we will discuss in the following section, the proposed quantum blockchain is a public and permissionless blockchain which satisfies the following properties:

1) A Decentralized architecture,
2) A quantum network with a distributed ledger,
3) Each node in this quantum network possesses quantum capabilities such as quantum storage and quantum state preparation,
4) Shared common quantum database.

Though blockchain has a wide range of applications, for the sake of illustration, we will use the example of a blockchain that manages a digital currency called *qCoin*. New transactions are proposed by those nodes who wish to transfer their funds to another node and are authenticated by their digital signature. There are various proposed digital signatures. Unconditionally secure digital signatures, such as Toeplitz Group Signature [31] and others, have been proposed and meet the following criteria: Unforgeability, Transferability, and Non-Repudiation.

The operation of the blockchain consists of the creation of new transactions and the construction of blocks that aggregate them. The process involves signature, broadcast, verification, encryption, and linking. Each transaction includes the information about the sender, receiver, amount, timestamp, and a list of transaction records demonstrating that the sender has sufficient funds to complete the transaction. The information of blocks containing transactions is encoded in an $n-$ qubit state. The data are then encrypted using the Generalized Quantum Gram-Schmidt process and transmitted using quantum key distribution to other nodes in the blockchain network and are collected in the pool of unconfirmed transactions. Each node checks these unconfirmed transactions with respect to their local copy of the blockchain in order to validate the encoded information. Further validation is performed using voting protocol before the verifier adds the block to its blockchain.

We do not have to presume that every single node is trustworthy in our vote-based consensus. Instead, the system functions as long as a certain proportion of nodes are trustworthy [32].

## IV. PROTOCOL FOR QUANTUM BLOCKCHAIN

The structure of the Quantum Blockchain is given in the Algorithm (1) in the Method section. Here we provide a detailed version of our scheme. For a single transaction round, we will now go over each phase of the protocol in detail. Each round of the protocol consists of the following four phases - the Transaction phase, the Verification phase, Consensus, and the Block Linking phase.

### A. THE TRANSACTION PHASE

*Definition 1:* A **transaction** $T_x$ is defined as a tuple $(\mathfrak{S}, \mathfrak{R}, \mathfrak{A}, \mathfrak{T})$, where $\mathfrak{S}$ is the sender of this transaction; $\mathfrak{R}$ are the receivers of this transaction $= \{(r_1), \ldots, (r_m)\}$; $\mathfrak{A}$ is the amount of the qCoin to be transmitted. And, $\mathfrak{T}$ are the sources, which is a list of transactions $(T_1, \ldots, T_n)$ to be redeemed by $T_x$. [31]

#### 1) STEP 1: CREATION OF THE TRANSACTION

For this round, we consider a node, say Alice, wish to send *qCoin* to another node, say Bob. Because it is Alice who wants to transfer her *qCoin*, she will initiate a Transaction $T_x$ and sign it with her digital signature. Then she will encode the information of this transaction $T_x$ containing her digital signature, Bob's address, the amount to be transferred, and the sources of this fund. The encoded state can be denoted as:

$$V_i = T_x (\mathfrak{S}, \mathfrak{R}, \mathfrak{A}, \mathfrak{T})$$

#### 2) STEP 2: BROADCAST

Once Alice has performed the encoding, she broadcasts the encoded transaction $V_i$ to the entire network. We for the practicality of the protocol, unlike other models, assume that not everyone in the network heard it. Let's say Charlie receives the broadcast; he will proceed to the verification phase.

We will follow a procedure that is comparable to the original Bitcoin transaction to finish a transaction [9]. When Alice spends the fund she received from an earlier transaction (for example, a transaction that reports Alice receiving 5 *qCoin* from someone). She wants to send Bob, say 3 *qcoin*. There will be two transactions, one $T_1$ ($\mathfrak{A}, \mathfrak{B}, 3, \mathfrak{T}$) sending 3 *qCoin* to Bob and the other $T_2$ ($\mathfrak{A}, \mathfrak{A}, 2, \mathfrak{T}$) sending 2 *qCoin* to herself, and both will have unique encoding because the message is distinct, necessitating the use of two different state vectors.

### B. THE VERIFICATION PHASE

Once Charlie heard the broadcast, he verifies the authenticity of the transaction based on his blockchain and voting protocol.

#### 1) STEP 1: TRANSACTION VERIFICATION

A transaction $T_x$ is considered to be valid if and only if the following holds [31]:
1) $T_x$ is properly signed by its sender $\mathfrak{S}$,
2) The sender of $T_x$ is one of the receivers in each of its source transactions $\mathfrak{T}$,
3) The certification & signature of $T_x$ evaluates the protection of all its source to be true,
4) None of its source transactions $\mathfrak{T}$ has been redeemed before.

Once the transaction satisfies all the above conditions, it is collected in the Log (a collection of valid but unconfirmed transactions that must be agreed upon by a shared consensus) owned by Charlie.

#### 2) STEP 2: PRELIMINARY BLOCK CREATION

Subsequently, the unconfirmed transactions from the log are aggregated to form a preliminary block. The preliminary block after encoding can be written in an $n-$ dimensional state given by:

$$\tilde{V}_i = (V_1, \cdots, V_k)$$

where $V_1$ to $V_k$ are the collection of unconfirmed transactions (in the transaction pool).

We are abandoning the classical blockchain approach of having blocks produced by individual nodes called 'miners,' as it is susceptible to attacks by quantum computers. Otherwise, a miner will have full freedom to create seemingly valid transactions and put them in a block. Instead, we propose that blocks be created in a decentralized manner.

### C. CONSENSUS

Through the consensus mechanism, nodes reach a shared conclusion regarding the transaction encoded in the preliminary block. The consensus algorithms employed in blockchain technology can be categorized into two basic types [33]. The first is proof-based consensus algorithms, which we typically see in permissionless blockchains such as proof-of-work (PoW) employed in Bitcoin. The second is vote-based consensus algorithms, which are most commonly used in permissioned blockchains and may include the Byzantine fault tolerance algorithms.

Each round of the consensus consists of the following three steps - the proposing step, the voting step, and the decision step.

#### 1) STEP 1: THE PROPOSING STEP

After completing the Transaction Verification step from the Verification phase, the proposing peer sends the preliminary block proposal to the voting peers. The proposal is signed by the proposing peer being the sender, and all voting peers as receivers.

#### 2) STEP 2: THE VOTING STEP

The proposal is sent to all voting peers. Voting peers enter the voting phase, during which they exchange votes across the network. The proposal created by any proposing peer is not sent to all the members of the network. Instead, using a Quantum Random Number Generator, an arbitrary $r$ number of peers are selected as voting peers for each round and for a particular proposing peer. This has two advantages. First, if each proposing peer only selects a subset of nodes instead of all nodes, it makes the Voting phase computationally less demanding and much faster. Second, randomly choosing a subset of voting peers makes it secure against the proxy attack, where the network can be flooded with proxy nodes by untrusted peers.

Other voting protocols can also be used, but that is not the focus of this paper. This information-theoretically secure protocol allows us to achieve consensus in the network connected by a quantum channel, provided that the number of dishonest parties is less than some specified percentage. The protocol can be adjusted by increasing $r$ to tolerate more dishonest nodes.

#### 3) STEP 3: THE DECISION STEP
Votes for a block after the Voting phase are sent to the proposing peer. Suppose any individual transaction is not confirmed by any of the voting peers. In that case, a wrong transaction message is broadcasted to the network, which can be used to implement protocols like the Two-phase commit protocol [34]. And that preliminary block is dropped, and the process is repeated.

### D. BLOCK CREATION
Once the Voting step is completed with a positive result, the proposing peers proceed to create the block and link it to their existing blockchain.

#### 1) STEP 1: BLOCK CREATION & LINKING
The final blocks are created using the Generalized Quantum Gram-Schmidt orthogonalization. Because each block is added one by one and in order of their timestamp, let us consider Charlie already has the following blockchain in his memory:

$$\tilde{W}_1, \cdots, \tilde{W}_{i-1}$$

where $\tilde{W}_1$ to $\tilde{W}_{i-1}$ represents the block elements of his blockchain.

In order to add the $i$th block $\tilde{W}_i$ in this blockchain, he uses Generalized Gram-Schmidt orthogonalization to the preliminary block $\tilde{V}_i$. As this Generalized Quantum Gram-Schmidt uses dimensional lifting, additional dimensions, say $m$, will be added. There is no restriction on $m$, but it will provide the bond on the maximum size of the blockchain [28].

So after the orthogonalization the final blockchain that Charlie has is of the form:

$$\tilde{W}_1, \cdots, \tilde{W}_{i-1}, \tilde{W}_i$$

where the information of each block is encoded in a $k = m+n$ dimensional orthogonalized states and are in order of their timestamp. Also note that since these blocks $\tilde{W}_j$ for $1 \leq j \leq i$ are orthogonal, they constitutes the a basis set of a subspace of $\mathbb{C}^{n+m}$.

#### 2) STEP 2: ENCRYPTION
The blockchain that has been created so far is unencrypted and is on the computational basis. Now we will encrypt each block in the blockchain. The encryption scheme is applied to a part of each block and is unique to each node in the network.

The information of any block is divided into two parts disclosed part and the encrypted part. The Disclosed part, which is the first $n$ component of the state encoding the blocks, consists of the original information of the preliminary block, which now encodes the confirmed transactions. This part remains on the computational basis. After the Generalized Quantum Gram-Schmidt process, additional $m-$ dimensions were added. The encryption scheme is applied to this part of the encoded block.

#### 3) ENCRYPTION SCHEME
1) Each member of the network selects a secret number $\theta \in [0, \pi]$.
2) Based on their number $\theta$, they create a unitary basis change transformation $U_\theta$.
3) On each of their block the member applies a unitary transformation $I^{\otimes n} \otimes U_\theta^{\otimes m}$.

By applying encryption scheme the disclosed part of each block remains in computational basis, but the encrypted second part is now encoded in $(|0\rangle \pm e^{i\theta}|1\rangle)/(\sqrt{2})$. Because the $\theta$ is secret and unique to each member of the network, no other member has access to the encrypted information of the block. So, for all other users it is in an unknown state and thus protected by the no-cloning theorem.

## V. FORKING CONDITION AND THEIR SOLUTION
There is also the possibility of two proposing peer mining the different blocks at the same time, resulting in a fork. There will then be ambiguity as to which block is considered the valid one.

The orthogonalization process and consensus used in our protocol make it very easy to tackle significant forking conditions that can arise in a blockchain. Consider the following two major forking scenarios:

### A. FORKING DUE TO DOUBLE SPENDING
*Definition 2:* **Double-spending** *is a potential flaw in a digital cash scheme in which a single unit of cryptocurrency is spent simultaneously more than once. This results in a discrepancy between the transaction record and the available currency. The main reason for double spending is that classical digital currency can be easily replicated.* [35]

Double spending remains a risk in a classical blockchain. The likelihood of a secret block being inserted into the quantum blockchain is very slim because it has to be accepted and verified by the network of voting peers.

Consider Alice has one *qCoin* and tries to spend it twice in two distinct transactions. She could attempt to do this by sending the same *qCoin* to two separate recipients. Both of these transactions will subsequently be added to the pool of unconfirmed transactions, which already contains a large number of unconfirmed transactions. As transactions (requests to send the *qCoin*) are broadcast, they will arrive at each node at slightly different times. If two transactions attempt to spend the same *qCoin*, each node will consider the first transaction it receives to be valid and the other invalid. However, once different nodes become mismatched, the validation of true balances becomes nearly impossible. Such a problem can be resolved by the use of a consensus algorithm, which syncs the various nodes.

Suppose two different proposing peers pick both transactions at the same time and start creating a block. Now suppose if both of them received the voting confirmation at nearly the same time. When the block is confirmed, both will wait for confirmation on their transaction from the Voting Phase of other proposing peers. Whichever transaction passes Voting Phase, more number of proposing peers will be validated, and another transaction will be pulled out from the network.

It is worth noting that the encrypted blocks of each peer contain the original information of the transaction, which is publicly available to all other peers in the network. Other members can see what transaction each peer has added by using appropriate surjective partial isometry, which in this case is nothing but measuring the first *n* component of the encrypted block on the computational basis (see the Theorem (1) for details).

### B. FORKING DUE TO BROADCAST

As for the practicality, we have assumed in our model not everyone hears the broadcast of every transaction or at least not at the same time. So it might happen that one peer hears one transaction before the next, while another hears the transaction in a different order. This is different from the Double-Spending case as here the same *qCoin* is not spent twice and all the transactions are valid. The following theorem resolves any such issues.

*Theorem 2:* Suppose one node in the network carries a different blockchain than others. Then the node can transform its local blockchain to the majority chain using a local unitary operation, provided all the transactions are valid.

The proof of the above theorem is given in Appendix B. So, as long as transactions are legitimate, no one is obligated to work in the same common chain at all times since they can turn their chain into a network-wide unique chain at any moment.

### VI. SECURITY ANALYSIS

The quantum blockchain scheme presented here is secured by the non-democratic nature of the Generalized Gram-Schmidt orthogonalization and quantum no-cloning theorem.

The following two theorems formally state and provide proof of the above argument. The first theorem guarantees the security of the blockchain, in the sense that no transaction can be modified without changing the final state of the blockchain. The second theorem is about the correctness, which states that the final state of the blockchain is unique for a given set of transactions.

*Theorem 3 (Security):* If any of the previous transactions are altered, then the final state of the blockchain will be changed.

*Proof:* Recall when the preliminary block $\tilde{V}_i = (V_1, \cdots, V_k)$ is encoded using Generalized Gram-Schmidt orthogonalization, we get the final block $\tilde{W}_i$. But if we consider a different set of transactions for the preliminary block then the final orthogonalized state will be different.

Consider on the contrary that a transaction in the *i*th block is changed without affecting the final state of the blockchain. So the initial encoding of *i*th block is changed from $\tilde{V}_i$ to say $\tilde{X}$. Then we claim that the final state remains the same after applying generalized Gram-Schmidt orthogonalization using the transformation operator $\mathcal{M}$ on this state. That is,

$$\mathcal{M}\left(\tilde{V}_1, \cdots, \tilde{V}_{i-1}, \tilde{X}, \cdots \tilde{V}_n\right) = \tilde{W}_1, \cdots, \tilde{W}_n \quad (5)$$

But from the Theorem (1) it follows that there exists a surjective partial isometry, namely the orthogonal projection operator $P$ such that it acts on the final state to give the initial state. Hence it follows that $P\tilde{W}_j = \tilde{V}_j$ for all $j = 1, \ldots, n$. But it also follows from the same argument that $P\tilde{W}_i = \tilde{X}$. Comparing the action of $P$ on the *i*th block we have that $\tilde{V}_i = X$. This is contradictory as we have assumed that the encoding of *i*th block is changed.

Hence we conclude that none of the previous transactions can be changed without altering the final state of the blockchain. Further, because the states of all the blocks in the final blockchain are in the orthogonal state, it is easy to distinguish between them. □

Further, if someone tries to alter any transaction from the *i*th block of Charlie. Then, it immediately invalidates all the successive blocks as they were orthogonalized with respect to the ordered set containing an encoding of the *i*th block. This means not only the final block $\tilde{W}_i$ is changed, but all the successive blocks i.e., $\tilde{W}_j$ for $j \geq i$ are also changed and hence invalidated. Suppose some peer in the network, say Charlie, has a different transaction encoded in any of their blocks then by measuring the disclosed part of the block on the computational basis, state information can be retrieved. In that case, all the other peers can know that there is an invalid transaction in the local blockchain owned by Charlie.

In comparison to the standard blockchain protocol where only the hash of the previous block is encoded in the next block by the linearity of the chain. In our case in some sense the information from all prior blocks that have appeared before the concerned block are encoded. This is because altering any state immediately invalidates all the encoded blocks that came after it.

*Theorem 4 (Correctness):* If $|\tilde{W}_1\rangle, \ldots, |\tilde{W}_m\rangle$ are obtained after performing the generalized Gram-Schmidt orthogonalization on the states $|\tilde{V}_1\rangle, \ldots, |\tilde{V}_m\rangle$, all in $\mathbb{C}^k$, where $k = n + m$. Then these vectors form a unique system of orthogonal states such that the following conditions are satisfied:

1) For all $1 \leq i \leq m$, the orthogonal projection $P$ of $\mathbb{C}^k$ onto $\mathbb{C}^n$ (as defined in equation (4)) sends $|\tilde{W}_i\rangle$ to $|\tilde{V}_i\rangle$.
2) The orthogonal projection of $\mathbb{C}^k$ onto $\mathbb{C}^m$ sends $|\tilde{W}_1\rangle, \ldots, |\tilde{W}_m\rangle$ to some (ordered) basis of the subspace $\mathbb{C}^m$. Applying the Gram-Schmidt process to this (ordered) basis gives the computational basis $|\mathbf{e}_1\rangle, \ldots, |\mathbf{e}_m\rangle$.

The proof and details of the above theorem are given in Ref. [28].

Furthermore, as discussed previously, the quantum no-cloning theorem makes the proposed blockchain even more robust than the standard classical blockchain. For a standard vote-based consensus algorithm, the security is provided by algorithms like Byzantine fault tolerance algorithms [36]. Hence if any hacker wants to change a particular transaction, he needs to hack into at least one-third of the network at the same time. There is a possibility that such attacks are feasible and can be aided by using proxies. But because of the encryption scheme that we are using in this blockchain protocol, it is not possible for any hacker to hack into even a single node of the network.

Further, as mentioned in Ref. [32], in earlier models that were proposed, the database is still somewhat vulnerable while it is locally stored. A possible attack scenario is when a malicious party equipped with a quantum computer works offline to forge the blockchain. To make the forged version appear legitimate, it modifies one of the previous transaction records to its profit and runs a Grover search [37] for a variant of other transactions within the same block such that its hash remains the same. When the search is complete, it hacks into all or some network nodes and replaces the legitimate database with the counterfeit version. But in this case, it is not possible for any hacker to change any of the past transactions because of the encrypted part of the block. Because the hacker does not know the value of $\theta$, which is unique for each member in the network and hence he does not know the quantum states of the encrypted part. Therefore, the quantum no-cloning theorem prevents such kind of attack. Even if he tries to alter the state to any arbitrary state, from the *security theorem* (3) it follows that the block encoding will be invalidated. Only the disclosed part of the block encoding is public. But changing the disclosed part without proper change in the encrypted part, again due to the *security theorem* (3), will make the block encoding invalid and also all the blocks that follows.

To sum up, the security of the proposed quantum blockchain protocol is ensured by two properties. Firstly, quantum mechanics principles such as the quantum no-cloning theorem and the Heisenberg uncertainty principle provide protection against counterfeiting. Secondly, it is secured by the non-democratic nature of Generalized Quantum Gram-Schmidt orthogonalization. As a result, even under quantum computing attacks, the proposed quantum blockchain scheme is secure. This is because the quantum capabilities of any hacker do not provide any advantage for this blockchain network.

## VII. QUANTUM TOKEN BASED ON QUANTUM BLOCKCHAIN

In this section, based on the proposed quantum blockchain, we present the inner workings of a Quantum Token called *qToken*, a quantum currency with no central authority.

*Definition 3:* A **quantum token** is defined by a pair of classical and quantum states $t_i = (c_i, |q_i\rangle)$, where $i$ is a serial number and $c_i$ is a transaction record made of classical bits and transaction is done by using a token machine or passing token to another person. [38]

These classical bits $c_i$ and quantum bits (qubits) $|q_i\rangle$ given to tokens should be in one-to-one correspondence so that no one can duplicate them. Namely, they obey $c_i \neq c_j \iff |q_i\rangle \neq |q_j\rangle$ for all serial numbers $i, j$.

As is the case with the majority of quantum money schemes, the main idea is based on the no-cloning theorem, which states that an unknown quantum state cannot be copied. So quantum mechanics provides apparent advantages for quantum states to be regarded as a cryptocurrency. And this new type of token offers a further advantage over just quantum blockchain. For example, being a physical quantity, there is no double-spending scenario.

When one uses the token machine, it implements our quantum blockchain scheme and each such token machine acts as an automated peer. The orthogonalized state is encoded in the token and the disclosed part of the block is used as the classical state $c_i$ for that token. Hence knowing disclosed part provides information about the value associated with the token. It can also be used to indicate the person to which the token belongs.

Even though the disclosed part is visible to the public as the serial number but the quantum state encoded in the token is still encrypted. So whenever a token is inserted in a token machine, the machine recognizes the serial number, i.e., disclosed classical part and retrieves the information about the corresponding encrypted quantum state vector. Then it is measured in an appropriate basis to validate its authenticity and completes the transaction.

The internal workings of the device are not accessible to the user. The security of this quantum token is guaranteed by the same logic as our quantum blockchain. And hence it is secure against counterfeiting. If the *qToken* is not part of our quantum blockchain network, then the transaction will be objected. The no-cloning theorem, in particular, ensures that quantum tokens cannot be duplicated or altered, prohibiting counterfeiting.

This system is difficult to hack because even when partial data are exposed or hacked from the system. The Gram-Schmidt orthogonalization which is highly non-democratic, necessitates the knowledge of all vectors before orthogonalizing the next vector. So a hacker needs to hack all of the information in the system, starting with the first transaction, which can be prevented easily.

This quantum token based on the quantum blockchain scheme has an advantage over other quantum token models that work like quantum coins [39], [40] since it contains genuine information about the transaction rather than arbitrary states. As a result, each coin has a distinct worth instead of all coins having the same value. A cryptocurrency ATM based on the quantum blockchain can also be prepared by adding a passcode similar to a peer's private key in the machine, preventing any other peer from using any lost or stolen token without the private key. The most well-known classical example of such an ATM at the moment is the Bitcoin ATM [41].

## VIII. CONCLUSION

Recent advancements in quantum computing have piqued the curiosity of researchers and developers towards Distributed Ledger Technologies (DLTs) such as blockchain, which are vulnerable to quantum computation attacks. A new quantum-resistant blockchain architecture is needed. This paper provides a functional model in this direction by formulating a practically realizable model of a fully Quantum Blockchain using generalized quantum Gram-Schmidt orthogonalization with dimensional lifting. Furthermore, it allows us to develop alternative variants by incorporating other consensus strategies. We have given a comprehensive analysis of forking scenarios with their countermeasures along with proof of security and robustness. These research works will allow us to continually use blockchains in a wide range of applications, even in the face of quantum computing attacks. We have also shown in the last section how this technology can be used to formulate a new type of quantum coin in the form of tokens, which offers greater flexibility.

Our research brings us one step closer to the development of a practical quantum cryptographic scheme. Our scheme has reasonably low computational complexity and requires limited quantum capabilities like quantum storage and quantum state preparation. Therefore, it is much easier to put into practice in real-world circumstances in comparison to other proposed methods in the same direction. Nevertheless, implementing multiparty schemes like these requires a proper quantum network. Additionally, the complexity analysis of our subroutine provided in Ref. [29] and [30], demonstrates that we still need more capable quantum computers to create a practical quantum blockchain. We are witnessing tremendous interest and progress in this direction. It would be worthwhile to investigate how we can minimize the resource counts even further in order to implement it even in a small-scale quantum computer. In future research, it will also be essential to examine any security challenges in practical implementation, such as due to side-channel attacks. In the UAV context, it has already been shown how we can retain the anonymity and secrecy outlined in Ref. [42], [43], and [44], while overcoming hardware and physical security concerns that were not considered in the theoretical model.

This proposed quantum blockchain can be regarded as a conceptual design to provide scientific development of a practical and fully quantum blockchain. Finally, because of faster processing speed, lower resources, and safer transactions, the quantum blockchain has a clear advantage over the classical blockchain. Therefore, it can be used for a wide range of applications while maintaining transparency and integrity of transactions, even under attack using quantum computers.

## METHODS

**Algorithm 1:** Protocol for the Quantum Blockchain

**Sender:**
**Input** : Signature $\mathfrak{S}$, Receivers $\mathfrak{R} = \{(r_1), \ldots, (r_m)\}$, qCoin amount $\mathfrak{A}$ & sources $\mathfrak{T} = (T_1, \ldots, T_n)$.

$V_j \leftarrow encode\{T_x\,(\mathfrak{S}, \mathfrak{R}, \mathfrak{A}, \mathfrak{T})\}$
**Broadcast:** $V_j$
**Minors:**
**Receive** : $V_1 \cdots V_n$

$Tverification_i \leftarrow verify\{V_i\}$  // local verification
**if** *for* $1 \leq i \leq n$, $Tverification_i = TRUE$ **then**
   $\tilde{V} = encode\{V_1, \cdots, V_n\}$
   **Function** Consensus($V$, $r$, Decision):
      $r \leftarrow rand\{\text{all nodes}\}$  // randomly selecting
      Decision $\leftarrow vote\{V, r\}$
      **return** Decision

   **Def** Encryption($W$):
      $encrypt\{W\} = I^{\otimes n} \otimes U_\theta^{\otimes m}(W)$
      $W_{enc} \leftarrow encrypt\{W\}$
      **return** $W_{enc}$

   **if** *Decision* = *True* **then**
      $\tilde{W}_j \leftarrow \mathcal{M}(\tilde{W}_1, \cdots, \tilde{W}_{j-1}, \tilde{V})$
      $\tilde{W}_{j,enc} \leftarrow encrypt\{\tilde{W}_j\}$
      $BC_j \leftarrow link\{BC_{j-1}, \tilde{W}_{j,enc}\}$  // blockchain with $i$ blocks
   **else**
      Reject
**else**
   Reject
**Access:**
**Input** : $\tilde{W}_{i,enc}$

Transaction Information $\leftarrow P(\tilde{W}_{i,enc})$

Most of the results in this work are determined via theoretical calculations. The basic quantum blockchain protocol proposed above is summarized in the Algorithm 1.

## APPENDIX A
## EXAMPLE SHOWING CALCULATION OF GRAM-SCHMIDT PROCEDURE

Assume we have n sets of states similar to

$$\{|1_i\rangle \otimes |2_i\rangle \otimes |3_i\rangle \otimes \cdots \otimes |n_i\rangle \mid i = 1 \ldots N\}$$

where $i = 1, \ldots, N$ denotes the number of states in the sets, and $n$ is the number of tensor products of components from $\mathbb{C}^2$.

We are now seeking to orthogonalize this set of $N$ states. By writing these vectors in terms of $2^n \times 1$ matrix, we may use Gram-Schmidt decomposition (because it is tensor products of $\mathbb{C}^2$). In order to write the corresponding vector in $[\mathbb{C}^2]^{\otimes n} := \underbrace{\mathbb{C}^2 \otimes \cdots \otimes \mathbb{C}^2}_{n}$ given a vector $x \in \mathbb{C}^{2^n}$, we must first grasp the conventional interpretation of a state $|\phi\rangle$ represented as a vector $x = (x_1, \ldots, x_{2^n}) \in \mathbb{C}^{2^n}$ in a tensor product. Qubits are represented by these states, which implies that each replica of $\mathbb{C}^2$ has the same standardized basis $\{|0\rangle, |1\rangle\}$. The following basis of $[\mathbb{C}^2]^{\otimes n}$ is induced by the computational basis $\{|0\rangle, |1\rangle\}$ of $\mathbb{C}^2$:

$$\mathcal{B} = \{|b_1\rangle|b_1\rangle \cdots |b_n\rangle : b_i \in \{0, 1\} \text{ for } i = 1, \ldots, n\}$$

It is worth noting that this basis is arranged in lexicographical order. If $|\phi_i\rangle$ is the $i$th element of $\mathcal{B}$, then $x$ is the vector that corresponds to $|\phi\rangle$ if

$$|\phi\rangle = x_1|\phi_1\rangle + \cdots + x_n|\phi_n\rangle$$

Now, an intriguing implication of lexicographical sequencing is that the $i$th element $|\phi_i\rangle$ and the binary expression of the $i$ have a beautiful connection : if $i$ has digits in its $n$-digit binary representation $b_1, b_2, \ldots, b_n$ then

$$|\phi_i\rangle = |b_1\rangle|b_2\rangle \cdots |b_n\rangle$$

For example: with $n = 5$, the 5-digit binary representation of 11 is $01011_2$, which means that $|\phi_{11}\rangle$ (the 12th element of the induced basis on the tensor product) is

$$|\phi_{11}\rangle = |0\rangle|1\rangle|0\rangle|1\rangle|1\rangle$$

Note that the state $|\phi_i\rangle \in [\mathbb{C}^2]^{\otimes n}$ is sometimes abbreviated as $|i\rangle$ in the literary works, however this appears to be incompatible with our notation.

Consider the following vector as an example.

$$\frac{1}{2}(0, 1, 0, i, 0, -1, -i, 0) \in \mathbb{C}^{2^3}$$

The tensor product's associated element will be

$$|\phi\rangle = \frac{1}{2}|\phi_2\rangle + \frac{i}{2}|\phi_4\rangle - \frac{1}{2}|\phi_6\rangle - \frac{i}{2}|\phi_7\rangle$$
$$= \frac{1}{2}|0\rangle|0\rangle|1\rangle + \frac{i}{2}|0\rangle|1\rangle|1\rangle - \frac{1}{2}|1\rangle|0\rangle|1\rangle - \frac{i}{2}|1\rangle|1\rangle|0\rangle.$$

## APPENDIX B
## PROOF OF THEOREM 5.1

*Proof:* Say if Charlie notices that the network has a different chain, he then investigates which chain the majority of individuals own. Assuming he was in sync up to the first $i$ th block, after which transactions are in a different order due to difference in broadcast receiving.

Say the blockchain of Charlie is $C_1 = \{\tilde{W}_1, \cdots, \tilde{W}_i, \tilde{X}_{i+1}, \cdots, \tilde{X}_n\}$ and majority of the network holds the chain $C = \{\tilde{W}_1, \cdots, \tilde{W}_n\}$. Intuitively it is clear because both the chains are essentially bases of the subspace, as previously stated, such a unitary transaction always exists. We will first show the existence of an operator that transforms the states in $C_1$ to *respective* states in $C$. Then we will show that this operator is unitary.

To make the calculation simpler, we assume that all the states $\tilde{W}$s and $\tilde{X}$s are of unit length. We begin by selecting unit length state vectors $P = \{p_{n+1}, \cdots, p_k\}$ each in $\mathbb{C}^k$ such that $C_1 \cup P$ consists of $k$ orthogonal state vectors and therefore is an orthogonal basis of $\mathbb{C}^k$. The existence of $P$ follows from a version of the Basis Extension Theorem. By applying the same result to $C$ we find another set $Q = \{q_{n+1}, \cdots, q_k\}$ each in $\mathbb{C}^k$ such that $C_1 \cup Q$ forms another orthogonal basis of $\mathbb{C}^k$. Note that this extension is to make sure that the operator $\mathcal{O}$ is unitary.

Now we define an operator using the above notation as:

$$\mathcal{O} = \sum_{j=1}^{i} \tilde{W}_j \tilde{W}_j^T + \sum_{j=i+1}^{n} \tilde{W}_j \tilde{X}_j^T + \sum_{j=n+1}^{k} q_j p_j^T \quad (6)$$

Now consider the following operations, using orthogonality and unity conditions we have:

$$\mathcal{O}\tilde{W}_l = \sum_{j=1}^{i} \tilde{W}_j \tilde{W}_j^T \tilde{W}_l + \sum_{j=i+1}^{n} \tilde{W}_j \tilde{X}_j^T \tilde{W}_l + \sum_{j=n+1}^{k} q_j p_j^T \tilde{W}_l$$
$$= \tilde{W}_l \tilde{W}_l^T \tilde{W}_l = \tilde{W}_l \cdot 1 = \tilde{W}_l$$

$$\mathcal{O}\tilde{X}_l = \sum_{j=1}^{i} \tilde{W}_j \tilde{W}_j^T \tilde{X}_l + \sum_{j=i+1}^{n} \tilde{W}_j \tilde{X}_j^T \tilde{X}_l + \sum_{j=n+1}^{k} q_j p_j^T \tilde{X}_l$$
$$= \tilde{W}_l \tilde{X}_l^T \tilde{X}_l = \tilde{W}_l \cdot 1 = \tilde{W}_l$$

$$\mathcal{O}p_l = \sum_{j=1}^{i} \tilde{W}_j \tilde{W}_j^T p_l + \sum_{j=i+1}^{n} \tilde{W}_j \tilde{X}_j^T p_l + \sum_{j=n+1}^{k} q_j p_j^T p_l$$
$$= q_l p_l^T p_l = q_l \cdot 1 = q_l$$

From the first two equations, the existence of the operator $\mathcal{O}$ is proved.

Now in order to prove that the operator $\mathcal{O}$ is unitary, consider an arbitrary state $v \in \mathbb{C}^k$. Since the set $C_1 \cup P$ forms the basis of $\mathbb{C}^k$, we can express the state $v$ as:

$$v = \sum_{j=1}^{i} \tilde{X}_j \tilde{W}_j + \sum_{j=i+1}^{n} \tilde{X}_j \tilde{X}_j + \sum_{j=n+1}^{k} \tilde{X}_j p_j \quad (7)$$

Now again using the orthogonality and unity conditions, we have the following equations

$$||v||_2^2 = ||\sum_{j=1}^{i} \tilde{X}_j \tilde{W}_j + \sum_{j=i+1}^{n} \tilde{X}_j \tilde{X}_j + \sum_{j=n+1}^{k} \tilde{X}_j p_j||$$

$$= \sum_{j=1}^{i} ||\tilde{X}_j \tilde{W}_j||_2^2 + \sum_{j=i+1}^{n} ||\tilde{X}_j \tilde{X}_j||_2^2 + \sum_{j=n+1}^{k} ||\tilde{X}_j p_j||_2^2$$

$$= \sum_{j=1}^{i} \tilde{X}_j^2 + \sum_{j=i+1}^{n} \tilde{X}_j^2 + \sum_{j=n+1}^{k} \tilde{X}_j^2 \qquad (8)$$

$$||\mathcal{O}v||_2^2 = ||\sum_{j=1}^{i} \tilde{X}_j \mathcal{O}\tilde{W}_j + \sum_{j=i+1}^{n} \tilde{X}_j \mathcal{O}\tilde{X}_j + \sum_{j=n+1}^{k} \tilde{X}_j \mathcal{O}p_j||_2^2$$

$$= ||\sum_{j=1}^{i} \tilde{X}_j \tilde{W}_j + \sum_{j=i+1}^{n} \tilde{X}_j \tilde{W}_j + \sum_{j=n+1}^{k} \tilde{X}_j q_j||_2^2$$

$$= \sum_{j=1}^{i} ||\tilde{X}_j \tilde{W}_j||_2^2 + \sum_{j=i+1}^{n} ||\tilde{X}_j \tilde{W}_j||_2^2 + \sum_{j=n+1}^{k} ||\tilde{X}_j q_j||_2^2$$

$$= \sum_{j=1}^{i} \tilde{X}_j^2 + \sum_{j=i+1}^{n} \tilde{X}_j^2 + \sum_{j=n+1}^{k} \tilde{X}_j^2 \qquad (9)$$

From the last two equations it follows that $||\mathcal{O}v||_2 = ||v||_2$ for all $v \in \mathbb{C}^k$. Hence it is proved that $\mathcal{O}$ is a unitary operator. □

## ACKNOWLEDGMENT
Author Contributions: Kumar Nilesh conceived and developed the conceptual and theoretical framework under the supervision of Prasanta K. Panigrahi. All results and proofs were produced by Kumar Nilesh Manuscript was written by Kumar Nilesh and proofread by Prasanta K. Panigrahi. Data Availability: The authors declare that all data supporting the findings of this study are available within the article or from the corresponding author upon reasonable request.

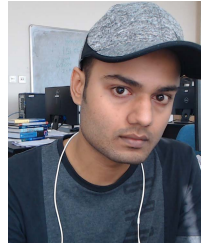

**KUMAR NILESH** received the integrated M.S. degree from the Department of Mathematics and Statistics, IISER Kolkata. He completed his master's thesis at the Indian Statistical Institute, Kolkata, which involved security proofs of Device-Independent QKD. He has worked on projects, such as quantum blockchain and multiparty quantum key distributions, among others.

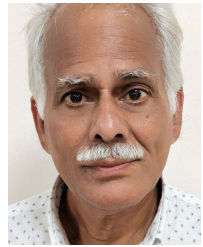

**PRASANTA K. PANIGRAHI** received the Ph.D. degree in quantum field theory from the University of Rochester. He is currently a Professor at IISER Kolkata. His research interests include field theory, quantum computing, quantum information processing, nonlinear dynamics, complex systems, wavelet analysis, computational physics, and biomedical signal processing.

• • •